\documentstyle[12pt]{article}
\author{Yu.N.Bespalov}
\title{\mbox{Crossed modules}\\
       \mbox{and}\\
       \mbox{quantum groups}\\
       \mbox{in}\\
       \mbox{braided categories $\;II$.}}
\data{August 11, 1994}

\def\ltimes{\mathrel{\,\hbox{\vrule height 4.5pt}\!\times}}
\def\newfigure{\begin{picture}(0,0)\end{picture}\newpage}
\def\interskip{\bigskip}
\newtheorem{theorem}{\bf Theorem}[section]
\newtheorem{proposition}{\bf Propositon}[section]
\newtheorem{lemma}{\bf Lemma}[section]
\newtheorem{definition}{\bf Definition}[section]
\newtheorem{remark}{\bf Remark}[section]

\newenvironment{proof}{\par\noindent{\bf Proof.}}{$\quad
\hbox{\vrule height 7pt width 0.5pt depth 1pt}\!
      \lower -2pt\hbox{${\vbox {\hrule width 8pt height 0.5pt depth 0pt}\atop
      \vbox {\hrule width 8pt height 0.5pt depth 0pt}}$}
      \!\hbox{\vrule height 7pt width 0.5pt depth 1pt}$\interskip}
\def\krr{\kern -.16667em}%
\def\kr{}%
\def\krrr{\kern -.3\unitlength}%
\newlength{\textwd}%
\def\hhstep{\kr\kr
\kern -.5\unitlength}
\def\hstep{\kr\kr
\kern .5\unitlength}
\def\step{\kr\kr
\kern \unitlength}
\def\Step{\kr\kr
\kern 2\unitlength}
\def\vvbox#1{{\offinterlineskip\vcenter{%
\def\coev{\kr
\begin{picture}(2,2)\put(1,0){\oval(2,2)[t]}\end{picture}}
\def\ev{\kr
\begin{picture}(2,2)\put(1,2){\oval(2,2)[b]}\end{picture}}
\def\hcoev{\kr
\begin{picture}(1,2)\put(.5,0){\oval(1,1)[t]}\end{picture}}
\def\hev{\kr
\begin{picture}(1,2)\put(.5,2){\oval(1,1)[b]}\end{picture}}
\def\COEV{\kr
\begin{picture}(2,2)\put(3,0){\oval(6,6)[t]}\end{picture}}
\def\EV{\kr
\begin{picture}(2,2)\put(3,2){\oval(6,6)[b]}\end{picture}}
\def\unit{\kr
\begin{picture}(0,2)
\put(0,0){\line(0,1){1}}\put(0,1.2){\circle{0.4}}
\end{picture}}
\def\counit{\kr
\begin{picture}(0,2)
\put(0,1){\line(0,1){1}}\put(0,.8){\circle{0.4}}
\end{picture}}
\def\Q##1{\kr
\begin{picture}(0,2)
\put(0,0){\line(0,1){0.4}}\put(0,1){\circle{1.2}}
\put(-0.6,0.4){\makebox(1.2,1.2)[cc]{$\scriptstyle ##1$}}
\end{picture}}
\def\O##1{\kr
\begin{picture}(0,2)
\put(0,0){\line(0,1){0.4}}\put(0,1.6){\line(0,1){0.4}}\put(0,1){\circle{1.2}}
\put(-0.6,0.4){\makebox(1.2,1.2)[cc]{$\scriptstyle ##1$}}
\end{picture}}
\def\S{\O{S}}                 \def\SS{\O{S^{-1}}}
\def\tS{\O{\widetilde S}}     \def\tSS{\O{\widetilde S^{-1}}}
\def\x{\kr
\begin{picture}(2,2)
\put(0,2){\line(1,-1){2}}\put(0,0){\line(1,1){.7}}\put(2,2){\line(-1,-1){.7}}
\end{picture}}
\def\xx{\kr
\begin{picture}(2,2)
\put(0,2){\line(1,-1){.7}}\put(0,0){\line(1,1){2}}\put(2,0){\line(-1,1){.7}}
\end{picture}}
\def\hx{\kr
\begin{picture}(1,2)
\put(0,2){\line(1,-2){1}}\put(0,0){\line(1,2){.35}}\put(1,2){\line(-1,-2){.35}}
\end{picture}}
\def\hxx{\kr
\begin{picture}(1,2)
\put(0,2){\line(1,-2){.35}}\put(0,0){\line(1,2){1}}\put(1,0){\line(-1,2){.35}}
\end{picture}}
\def\d{\kr
\begin{picture}(1,2)\put(0,2){\line(1,-2){1}}\end{picture}}
\def\dd{\kr
\begin{picture}(1,2)\put(0,0){\line(1,2){1}}\end{picture}}
\def\hd{\kr
\begin{picture}(1,2)
\put(0,2){\line(1,-2){.5}}
\put(.5,1){\line(0,-1){1}}
\end{picture}}
\def\hdd{\kr
\begin{picture}(1,2)
\put(1,2){\line(-1,-2){.5}}
\put(0,1){\line(0,-1){1}}
\end{picture}}
\def\ld{\kr
\begin{picture}(1,2)
\put(1,0){\oval(2,2)[lt]}\put(1,0){\line(0,1)2}
\end{picture}}
\def\Ld{\kr
\begin{picture}(2,2)
\put(2,0){\oval(4,2)[lt]}\put(2,0){\line(0,1)2}
\end{picture}}
\def\cd{\kr
\begin{picture}(2,2)
\put(1,0){\oval(2,2)[ct]}\put(1,1){\line(0,1)1}
\end{picture}}
\def\hdcd{\kr
\begin{picture}(1,2)
\put(0,2){\line(1,-2){.5}}
\put(.5,0){\oval(1,1)[ct]}\put(.5,.5){\line(0,1){.5}}
\end{picture}}
\def\hddcd{\kr
\begin{picture}(1,2)
\put(1,2){\line(-1,-2){.5}}
\put(.5,0){\oval(1,1)[ct]}\put(.5,.5){\line(0,1){.5}}
\end{picture}}
\def\hcd{\kr
\begin{picture}(1,2)
\put(.5,0){\oval(1,1)[ct]}\put(.5,.5){\line(0,1){1.5}}
\end{picture}}
\def\Cd{\kr
\begin{picture}(4,2)
\put(2,0){\oval(4,2)[ct]}\put(2,1){\line(0,1)1}
\end{picture}}
\def\rd{\kr
\begin{picture}(1,2)
\put(0,0){\oval(2,2)[rt]}\put(0,0){\line(0,1)2}
\end{picture}}
\def\Rd{\kr
\begin{picture}(2,2)
\put(0,0){\oval(4,2)[rt]}\put(0,0){\line(0,1)2}
\end{picture}}
\def\lu{\kr
\begin{picture}(1,2)
\put(1,2){\oval(2,2)[lb]}\put(1,0){\line(0,1)2}
\end{picture}}
\def\Lu{\kr
\begin{picture}(2,2)
\put(2,2){\oval(4,2)[lb]}\put(2,0){\line(0,1)2}
\end{picture}}
\def\cu{\kr
\begin{picture}(2,2)
\put(1,2){\oval(2,2)[cb]}\put(1,0){\line(0,1)1}
\end{picture}}
\def\hcu{\kr
\begin{picture}(1,2)
\put(.5,2){\oval(1,1)[cb]}\put(.5,0){\line(0,1){1.5}}
\end{picture}}
\def\Cu{\kr
\begin{picture}(4,2)
\put(2,2){\oval(4,2)[cb]}\put(1,0){\line(0,1)1}
\end{picture}}
\def\ru{\kr
\begin{picture}(1,2)
\put(0,2){\oval(2,2)[rb]}\put(0,0){\line(0,1)2}
\end{picture}}
\def\Ru{\kr
\begin{picture}(2,2)
\put(0,2){\oval(4,2)[rb]}\put(0,0){\line(0,1)2}
\end{picture}}
\def\k{\kr
\begin{picture}(1,2)
\put(0,2){\oval(2,1)[rb]}
\put(0,0){\oval(2,1)[rt]}
\put(0,0){\line(0,1)2}
\end{picture}}
\def\ro##1{\kr
\begin{picture}(2,2)
\put(.4,0){\oval(.8,.8)[lt]}\put(1.6,0){\oval(.8,.8)[rt]}
\put(1,0.4){\circle{1.2}}
\put(0.4,-0.2){\makebox(1.2,1.2)[cc]{$\scriptstyle ##1$}}%
\end{picture}}
\def\Ro##1{\kr
\begin{picture}(4,2)
\put(1.4,0){\oval(2.8,1.2)[lt]}\put(2.6,0){\oval(2.8,1.2)[rt]}
\put(2,.6){\circle{1.2}}
\put(1.4,0){\makebox(1.2,1.2)[cc]{$\scriptstyle ##1$}}%
\end{picture}}
\def\r{\ro{\cal R}}           \def\rr{\ro{{\cal R}^{-1}}}
\def\ra{\ro{{\cal R}_A}}        \def\rra{\ro{{\cal R}^{-1}_A}}
\def\rb{\ro{{\cal R}_B}}        \def\rrb{\ro{{\cal R}^{-1}_B}}
\def\rh{\ro{{\cal R}_H}}
\def\R{\Ro{\cal R}}           \def\RR{\Ro{{\cal R}^{-1}}}
\def\Ra{\Ro{{\cal R}_A}}        \def\RRa{\Ro{{\cal R}^{-1}_A}}
\def\Rb{\Ro{{\cal R}_B}}        \def\RRb{\Ro{{\cal R}^{-1}_B}}
\def\Rh{\Ro{{\cal R}_H}}
\def\id{\kr
\begin{picture}(0,2)\put(0,0){\line(0,1)2}\end{picture}}
\def\obj##1{\settowidth{\textwd}{$##1$}%
\raise .2\unitlength\hbox{\kern -.5\textwd $##1$ \kern -.5\textwd \krrr}}
\def\Obj##1{\settowidth{\textwd}{$##1$}%
\raise 1.1\unitlength\hbox{\kern -1\textwd $##1$}}
\def\hhbox##1{\hbox{%
\kern -4.45\unitlength
\def\coev{\kr
\begin{picture}(1,1)\put(.5,0){\oval(1,1)[t]}\end{picture}}
\def\ev{\kr
\begin{picture}(1,1)\put(.5,1){\oval(1,1)[b]}\end{picture}}
\def\ld{\kr
\begin{picture}(1,1)
\put(1,0){\oval(2,2)[lt]}\put(1,0){\line(0,1)1}
\end{picture}}
\def\Ld{\kr
\begin{picture}(2,1)
\put(2,0){\oval(4,2)[lt]}\put(2,0){\line(0,1)1}
\end{picture}}
\def\rd{\kr
\begin{picture}(1,1)
\put(0,0){\oval(2,2)[rt]}\put(0,0){\line(0,1)1}
\end{picture}}
\def\Rd{\kr
\begin{picture}(2,1)
\put(0,0){\oval(4,2)[rt]}\put(0,0){\line(0,1)1}
\end{picture}}
\def\cd{\kr
\begin{picture}(1,1)
\put(.5,0){\oval(1,1)[ct]}\put(.5,.5){\line(0,1){.5}}
\end{picture}}
\def\lu{\kr
\begin{picture}(1,1)
\put(1,1){\oval(2,2)[lb]}\put(1,0){\line(0,1)1}
\end{picture}}
\def\Lu{\kr
\begin{picture}(2,1)
\put(2,1){\oval(4,2)[lb]}\put(2,0){\line(0,1)1}
\end{picture}}
\def\cu{\kr
\begin{picture}(1,1)
\put(.5,1){\oval(1,1)[cb]}\put(.5,0){\line(0,1){.5}}
\end{picture}}
\def\ru{\kr
\begin{picture}(1,1)
\put(0,1){\oval(2,2)[rb]}\put(0,0){\line(0,1)1}
\end{picture}}
\def\Ru{\kr
\begin{picture}(2,1)
\put(0,1){\oval(4,2)[rb]}\put(0,0){\line(0,1)1}
\end{picture}}
\def\hru{\kr
\begin{picture}(.5,1)
\put(0,1){\oval(1,1)[rb]}\put(0,0){\line(0,1)1}
\end{picture}}
\def\hrd{\kr
\begin{picture}(.5,1)
\put(0,0){\oval(1,1)[rt]}\put(0,0){\line(0,1)1}
\end{picture}}
\def\id{\kr
\begin{picture}(0,1)\put(0,0){\line(0,1)1}\end{picture}}
\def\d{\kr
\begin{picture}(.5,1)\put(0,1){\line(1,-2){0.5}}\end{picture}}
\def\dd{\kr
\begin{picture}(.5,1)\put(0,0){\line(1,2){0.5}}\end{picture}}
##1}}
#1}\normalbaselines}}
\def\object#1{\settowidth{\textwd}{$#1$}%
                        \hbox{%
                        \kern -.5\textwd $#1$ \kern -.5\textwd}}
\def\map#1#2#3{\vcenter{\hbox{$#2\;$}}
                     \vcenter{\settowidth{\textwd}{$#1$}
	                      \hbox{\kern -.5\textwd $#1$ \kern -.5\textwd}
			      \hbox{\begin{picture}(0,2)
                                          \put(0,2){\vector(0,-1)2}
                                    \end{picture}}
                              \settowidth{\textwd}{$#3$}
	                      \hbox{\kern -.5\textwd $#3$ \kern -.5\textwd}}}

\begin{document}

\maketitle

\begin{abstract}
This is the second part of the paper.
Results of the first part about crossed modules are applied here to study
of quantum groups in braided categories.
Correct cross product in the class of quantum braided groups is built.
Criterion when quantum braided group is cross product is otained.
\end{abstract}

\section{Quantum braided groups.}

\begin{definition}(\cite{Majid8})
Bialgebra (Hopf algebra) $H$ in braided category $\cal C$ is
{\em quasitriangular} (or {\em quantum braided group} in the second case)
if it has
\begin{itemize}
\item
other bialgebra (Hopf algebra) structure with the same algebra and counit
and with the second comultiplication $\widetilde\Delta$
(and antipode $\widetilde S$);
\item
morphisms
${\cal R}\,,{\cal R}^{-1}:\,\underline 1\rightarrow H\otimes H$
("elements" of $H\otimes H$) such that:
\begin{equation}
{\cal R}\cdot{\cal R}^{-1}
\quad:=\quad
\vvbox{\hbox{\r\step\rr}
       \hbox{\id\Step\hx\Step\id}
       \hbox{\cu\step\cu}}
\quad ={}\quad
\vvbox{\hbox{\unit\step\unit}
       \hbox{\id\step\id}}
\quad ={}\quad
\vvbox{\hbox{\rr\step\r}
       \hbox{\id\Step\hx\Step\id}
       \hbox{\cu\step\cu}}
\quad=:\quad
{\cal R}^{-1}\cdot{\cal R}
\label{invertible}
\end{equation}
\begin{equation}
\matrix{
({\rm id}\otimes\Delta){\cal R}
\quad:=\quad
\vvbox{\hbox{\r}
       \hbox{\id\step\hstep\hcd}}
\quad ={}\quad
\vvbox{\hbox{\R}
       \hbox{\id\step\r\step\id}
       \hhbox{\krrr\cu\Step\id\step\id}}
\quad=:\quad
{\cal R}_{13}\cdot{\cal R}_{12}
\cr
(\widetilde\Delta\otimes{\rm id}){\cal R}
\quad:=\quad
\vvbox{\hbox{\hstep\r}
       \hbox{\hcd\obj{\widetilde\Delta}\hstep\step\id}}
\quad ={}\quad
\vvbox{\hbox{\R}
       \hbox{\id\step\r\step\id}
       \hhbox{\krrr\id\step\id\Step\cu}}
\quad=:\quad
{\cal R}_{23}\cdot{\cal R}_{13}
}
\label{rmatrix}
\end{equation}
\begin{equation}
\widetilde\Delta\cdot{\cal R}
\quad:=\quad
\vvbox{\hbox{\cd\Obj{\widetilde\Delta}\step\r}
       \hbox{\id\Step\hx\Step\id}
       \hbox{\cu\step\cu}}
\quad ={}\quad
\vvbox{\hbox{\r\step\cd}
       \hbox{\id\Step\hx\Step\id}
       \hbox{\cu\step\cu}}
\quad=:\quad
{\cal R}\cdot\Delta
\label{adjoint}
\end{equation}
\end{itemize}
\end{definition}

\begin{proposition}
Let $H$ be quasitriangular braided Hopf algebra.
Then the following identitieis are true:
\begin{equation}
\vvbox{\hbox{\r}
       \hbox{\counit\Step\id}}
\quad ={}\quad
\vvbox{\hbox{\unit}}
\quad ={}\quad
\vvbox{\hbox{\r}
       \hbox{\id\Step\counit}}
\qquad\qquad
\vvbox{\hbox{\rr}
       \hbox{\id\Step\id}}
\quad ={}\quad
\vvbox{\hbox{\r}
       \hbox{\id\Step\SS}}
\quad ={}\quad
\vvbox{\hbox{\r}
       \hbox{\tSS\Step\id}}
\end{equation}
\end{proposition}

This is analog of the Drinfel'd's definition \cite{Drinfeld1} but without
assumption that $\widetilde\Delta=\Delta^{\rm op}$.
In general case $H_{\rm op}$ with new comultiplication $\Delta^{\rm op}$
is Hopf algebra in $\overline{\cal C}$ but not in $\cal C$
(proposition \ref{opHopf}).
So it is unnatural to compare $\widetilde\Delta$ and $\Delta^{\rm op}$.
Majid suggested to consider quantum braided group $(H,{\cal R})$ together
with a suitable class of modules over $H$ for which $\widetilde\Delta$ is
in some sence 'opposite' to $\Delta$.
(We consider right modules instead of left ones.)

\begin{definition}(\cite{Majid8})
Let $H$ be quasitriangular braided Hopf algebra and $X$ is right module over
$H$.
Comultiplication $\widetilde\Delta$ is {\em opposite with respect to} $X$ if
\begin{equation}
\matrix{\object{X}\step\hstep\object{H}\hstep\cr
        \vvbox{\hhbox{\krrr\id\step\cd\obj{\widetilde\Delta}}
               \hbox{\ru\step\id}
               \hbox{\xx}}\cr
	\object{H}\Step\object{X}}
\quad ={}\quad
\vvbox{\hhbox{\krrr\id\step\cd}
       \hbox{\hx\step\id}
       \hbox{\id\step\ru}}
\label{opposite}
\end{equation}
Denote by ${\cal C}_H^{\cal R}$ the full subcategory of ${\cal C}_H$
with objects $X$ such that $\widetilde\Delta$ is opposite with respect to $X$.
\end{definition}

It is convenient to describe the category ${\cal C}^{\cal R}_H$ in terms of
crossed modules.
For right $H$-module $X$ denote by $X^{\cal R}$ object with additional
comodule structure
\begin{equation}
\vvbox{\hbox{\id\step\r}
       \hbox{\ru\Step\id}}
\label{comodstr}
\end{equation}

\begin{proposition}
Let $Y$ is right module over quasitriangular braided Hopf algebra $H$.
Then $\widetilde\Delta$ is opposite with respect to $Y$ iff $Y^{\cal R}$
is crossed module.

In this case for any module $X\;$:
\begin{equation}
(X\otimes Y)^{\cal R}=X^{\cal R}\otimes Y^{\cal R}
\end{equation}
\end{proposition}

\begin{proof} See figures {\ref{brm}}-{\ref{brmm}}
\end{proof}

So one can identify ${\cal C}_H^{\cal R}$ with full subcategory of
${\cal C}_H^H$.

\begin{proposition}
\label{qugrmod}
${\cal C}_H^{\cal R}$ is monoidal subcategory of ${\cal C}_H$ with braiding
\begin{equation}
\vvbox{\hbox{\id\step\id\step\r}
       \hhbox{\krrr\id\step\ru\Step\id}
       \hbox{\hx\Step\step\id}
       \hbox{\id\step\d\step\dd}
       \hbox{\id\Step\ru}}
\label{modulebraiding}
\end{equation}
\end{proposition}

\begin{proof}
Formulas for braiding can be obtained at once from (\ref{crossedmodulePsi}).
\end{proof}

\begin{lemma}
Let $H$ be quasitriangular Hopf algebra in $\cal C$ and $X$ module from
${\cal C}^{\cal R}_H$.
\begin{itemize}
\item
Then
\begin{equation}
\matrix{\object{H}\step\object{X}\cr
\vvbox{\hbox{\tS\step\id}
       \hbox{\hx}
       \hbox{\ru}}}
\quad ={}\quad
\matrix{\object{H}\step\object{X}\cr
\vvbox{\hbox{\SS\step\id}
       \hbox{\hxx}
       \hbox{\ru}}}
\label{changeantipode}
\end{equation}
\item
If there exist right dual $X^\vee$ in $\cal C$ then
\begin{equation}
\matrix{\object{X}\step\object{X^\vee}\step\object{H}\cr
\vvbox{\hbox{\id\step\id\step\S}
       \hbox{\id\step\hx}
       \hbox{\ru\step\id}
       \hbox{\ev}}}
\quad ={}\quad
\matrix{\object{X}\step\object{X^\vee}\step\object{H}\cr
\vvbox{\hbox{\id\step\id\step\tSS}
       \hbox{\id\step\hxx}
       \hbox{\ru\step\id}
       \hbox{\ev}}}
\label{changeantipode1}
\end{equation}
\end{itemize}
\end{lemma}

\begin{proof}
See figures {~\ref{proofchant}}, \ref{proofchantt}.
\end{proof}

\begin{proposition}
Let $X$ is right module from ${\cal C}_H^{\cal R}$ and there exists
left dual ${}^\vee X$ (resp. right dual $X^\vee$) then
$({}^\vee X)^{\cal R}={}^\vee (X^{\cal R})$
(resp. $(X^\vee )^{\cal R}=(X^{\cal R})^\vee$) $\quad{\rm and}\quad
{}^\vee X$ (resp. $X^\vee$) belongs to ${\cal C}_H^{\cal R}$.
\end{proposition}

\begin{proof}
Proposition is a direct corollary of the previous lemma and the identities
$({\rm id}\otimes S^{\pm 1})\circ{\cal R}=
 (\widetilde S^{\pm 1}\otimes{\rm id})\circ{\cal R}$.
For example, proof for the left dual is on the figure \ref{proofld}.
\end{proof}

\interskip
For any quantum group $(A,{\cal R}_A)$ in $\cal C$ one can define another
one $(A_{\widetilde{\rm op}},{\cal R}_{A_{\widetilde{\rm op}}})$
in category $\overline{\cal C}$ where $A_{\widetilde{\rm op}}$ is
Hopf algebra in $\overline{\cal C}$ with new comultiplication and
$\cal R$-element:
\begin{equation}
\widetilde\Delta^{\rm op}
\quad:=\quad
\vvbox{\hbox{\hcd\obj{\widetilde\Delta}}
       \hbox{\hxx}}
\qquad\qquad
{\cal R}_{A_{\widetilde{\rm op}}}\quad:=\quad
\vvbox{\hbox{\rra}
       \hbox{\xx}}
\end{equation}

\begin{lemma}
\label{opqugr}
$(A_{\widetilde{\rm op}},{\cal R}_{A_{\widetilde{\rm op}}})$ is
quantum group in category $\overline{\cal C}$.
Braided categories $\overline{{\cal C}^{{\cal R}_A}_A}$ and
$\overline{\cal C}
                  ^{{\cal R}_{A_{\widetilde{\rm op}}}}
                  _{A_{\widetilde{\rm op}}}$
coincide.
\end{lemma}

\interskip
The basic formulas for usual quantum groups \cite{Drinfeld2} has analogs
in our more general context.
But some of them exist only in a relative form: actions on arbitrary modules
from ${\cal C}^{\cal R}_H$ take part in them.
One can obtain habitual formulas for usual quantum groups if consider
action on unit element of regular module.

\begin{proposition}
(Relative form of the Yang-Baxter equation.)
If $X\in{\rm Obj}({\cal C}^{\cal R}_H)$ then
\begin{equation}
\matrix{\object{X}\Step\Step\step\cr
	\vvbox{\hbox{\id\step\r}
               \hbox{\hx\Step\id}
               \hbox{\id\step\Ru\r}
               \hbox{\id\step\Ru\Step\id}
               \hbox{\id\step\id\step\r\step\id}
               \hbox{\id\step\hxx\Step\hcu}
               \hbox{\hcu\step\id\Step\hstep\id}}}
\quad ={}\quad
\matrix{\object{X}\Step\Step\step\cr
	\vvbox{\hbox{\id\step\R}
               \hbox{\hx\step\r\step\id}
               \hbox{\id\step\ru\Step\hcu}
               \hbox{\id\step\id\step\r\hstep\id}
               \hbox{\id\step\hx\Step\id\hstep\id}
               \hbox{\hcu\step\Ru\hstep\id}}}
\label{YB}
\end{equation}
\end{proposition}

\begin{proof}
Identity (\ref{YB}) directly follows from (\ref{adjoint}), (\ref{rmatrix}).
\end{proof}

If we consider $X\in{\rm Obj}({\cal C}_H^{\cal R})$ as a crossed module
then the automorphism  $S^2$ of $X\in{\rm Obj}({\cal C})$
(see definition \ref{squareantipode}) is a result of action $\triangleleft u$,
where 'ellement' $u$ is defined in the following way:
\begin{equation}
\vvbox{\hbox{\Q{u}}}
\quad=\quad
\vvbox{\hbox{\r}
       \hbox{\id\Step\S}
       \hbox{\cu}}
\quad=\quad
\vvbox{\hbox{\r}
       \hbox{\tS\Step\id}
       \hbox{\cu}}
\label{udef1}
\end{equation}

\begin{proposition}
\begin{itemize}
\item
'Element' $u$ has inverse
\begin{equation}
\vvbox{\hbox{\Q{u^{-1}}}}
\quad=\quad
\vvbox{\hbox{\r}
       \hbox{\S\Step\SS}
       \hbox{\cu}}
\quad=\quad
\vvbox{\hbox{\r}
       \hbox{\tSS\Step\tS}
       \hbox{\cu}}
\label{udef2}
\end{equation}
and interweave antipodes in the following way
\begin{equation}
\vvbox{\hbox{\tS\Step\Q{u}}
	\hbox{\cu}}
\quad =\quad
\vvbox{\hbox{\Q{u}\Step\S}
	\hbox{\cu}}
\label{adjantipode}
\end{equation}
\item
The following (relative variant of the formula for $\Delta u$)
is true for any modules $X$ and $Y$ from ${\cal C}_H^{\cal R}$:
\begin{equation}
{}^{({\cal C}^{\cal R}_H)}\Psi_{Y,X}\circ(\triangleleft u)
                                  \circ{}^{({\cal C}^{\cal R}_H)}\Psi_{X,Y}=
{}^{\cal C}\Psi_{Y,X}\circ ((\triangleleft u)\otimes(\triangleleft u))
                                  \circ{}^{\cal C}\Psi_{X,Y}
\end{equation}
\end{itemize}
\end{proposition}

\begin{proof}
Proof of (\ref{adjantipode}) is on the figure ~\ref{proofadjant}.
Formula (\ref{udef2}) is the corollary of (\ref{adjantipode}).
The second part is a special case of (\ref{antipodetensor}).
\end{proof}

If one replace $(H,{\cal R})$ by
$(H_{\widetilde{\rm op}},{\cal R}_{\widetilde{\rm op}})$
then new 'elements'
\begin{equation}
S^{-1}u^{-1}=\widetilde S^{-1}u^{-1}=
\vvbox{\hbox{\r}
       \hbox{\id\Step\O{S^{-2}}}
       \hbox{\xx}
       \hbox{\cu}}
\enspace=\enspace
\vvbox{\hbox{\r}
       \hbox{\O{\widetilde S^{-2}}\Step\id}
       \hbox{\xx}
       \hbox{\cu}}
\qquad
S^{-1}u=\widetilde S^{-1}u=
\vvbox{\hbox{\r}
       \hbox{\xx}
       \hbox{\id\Step\SS}
       \hbox{\cu}}
\enspace=\enspace
\vvbox{\hbox{\r}
       \hbox{\xx}
       \hbox{\tSS\Step\id}
       \hbox{\cu}}
\end{equation}
are obtained instead of the $u$ and $u^{-1}$.
Our $S^{-1}u$ coincide with $u$ from \cite{Drinfeld2}, \cite{RT}.

\interskip
Futher results can be obtained if we suppose that category $\cal C$
is ribbon. Then one can defined ribbon structure on quantum braided group
$H$ in a such way that category ${\cal C}^{\cal R}_H$ is ribbon.
Ribbon structure is wellbehaved under cross products and transmutation
and so on. This results will be published anywhere.

\interskip
\begin{remark}
\begin{itemize}
\item
In definition of quasitriangular bialgebra $(H,{\cal R})$ coassociativity
and bialgebra axiom for second comultiplication $\widetilde\Delta$ are
corollaries of other axioms.
\item
In definition of quantum braided group one can replase assumption that
$\widetilde S$ exists by invertibility condition for $u$,
defined by the first identity from (\ref{udef1}).
\end{itemize}
\end{remark}

\begin{proof}
The proof of the first part:
It follows from (\ref{rmatrix}) that
\begin{equation}
\vvbox{\hbox{\rr}
       \hbox{\id\step\hstep\hcd}}
\quad ={}\quad
\vvbox{\hbox{\RR}
       \hbox{\id\step\rr\step\id}
       \hbox{\hxx\Step\id\step\id}
       \hhbox{\cu\Step\id\step\id}}
\qquad\qquad
\vvbox{\hbox{\hstep\rr}
       \hbox{\hcd\obj{\widetilde\Delta}\hstep\step\id}}
\quad =\quad
\vvbox{\hbox{\RR}
       \hbox{\id\step\rr\step\id}
       \hbox{\id\step\id\Step\hxx}
       \hhbox{\id\step\id\Step\cu}}
\label{rrmatrix}
\end{equation}
After that bialgebra axiom can be verifyed directly:
$$\widetilde\Delta\circ\cdot_H=
  {\cal R}(\Delta\circ\cdot_H){\cal R}^{-1}=
  {\cal R}(\cdot_{H\otimes H}\circ(\Delta\otimes\Delta)){\cal R}^{-1}=$$
$$ =\cdot_{H\otimes H}\circ
     ({\cal R}\Delta{\cal R}^{-1}\otimes{\cal R}\Delta{\cal R}^{-1})=
   \cdot_{H\otimes H}\circ(\widetilde\Delta\otimes\widetilde\Delta)     $$
Then one can use (\ref{rmatrix}), (\ref{rrmatrix}),
coassociativity of $\Delta$,
bialgebra axiom for $(H,\Delta)$ and $(H,\widetilde\Delta)$
to prove coassociativity of $\widetilde\Delta$:
$$(\widetilde\Delta\otimes{\rm id})\circ\widetilde\Delta=
  {\cal R}_{23}{\cal R}_{13}{\cal R}_{12}
    ((\Delta\otimes{\rm id})\circ\Delta)
    {\cal R}_{12}^{-1}{\cal R}_{13}^{-1}{\cal R}_{23}^{-1}=
 ({\rm id}\otimes\widetilde\Delta)\circ\widetilde\Delta       $$
\par
The sketch of the proof of the second part:
\par
It's easy to see that morphism $\widetilde S$ defined by (\ref{adjantipode})
is right antipode for $(H,\widetilde\Delta)$.
\par
{}From previous fact and (\ref{rmatrix}) one can obtain
$$\vvbox{\hbox{\r}
	 \hbox{\tS\Step\id}}
  \quad=\quad
  \vvbox{\hbox{\r}
	 \hbox{\id\Step\S}}    $$
$\widetilde S$ is invertible by definition. Then
$$\vvbox{\hbox{\r}
	 \hbox{\tSS\Step\id}}
  \quad=\quad
  \vvbox{\hbox{\r}
	 \hbox{\id\Step\SS}}    $$
\par
{}From definition of $\widetilde S$ and the last formule it follows that
$u^{-1}$ is the same as in (\ref{udef2}).
\par
$\widetilde S$ is left antipode for bialebra $(H,\widetilde\Delta)$.
\end{proof}

\begin{remark}
Objects of the category ${}_H^{\cal R}{\cal C}$ are left modules satisfying
the following condition:
\begin{equation}
\vvbox{\hbox{\hcd\obj{\widetilde\Delta}\step\id}
       \hbox{\id\step\hx}
       \hbox{\lu\step\id}}
\quad =\quad
\vvbox{\hbox{\hcd\step\id}
       \hbox{\id\step\lu}
       \hbox{\xx}}
\end{equation}
They became left crossed modules if one define coaction as the following:
\begin{equation}
\vvbox{\hbox{\r\step\id}
       \hbox{\id\Step\hx}
       \hbox{\Lu\step\id}
       \hbox{\Step\hx}}
\end{equation}
Category ${}_H^{\cal R}{\cal C}$ is braided with
\begin{equation}
{}^{({}_H^{\cal R}{\cal C)}}\Psi_{X,Y}:=
\matrix{\Step\step\object{X}\step\object{Y}\cr
        \vvbox{\hbox{\r\step\id\step\id}
	       \hbox{\id\Step\hx\step\id}
	       \hbox{\Lu\step\lu}
	       \hbox{\Step\x}}}
\end{equation}
\par
Braided categories ${}^{\cal R}_H{\cal C}$ and ${\cal C}^{\cal R}_H$ are
isomorphic: left module corresponding to the right one has the same underlying
object and action defined by the formula (\ref{changeantipode}).
So one can replace everywhere right modules by the left.
But formulas for the last case are more complicated.
\end{remark}

\section{Quantum group cross product.}

In this section we give cross product construction for quantum
groups in braided categories which generalize Majid's bosonization
\cite{Majid6} and prove analog of Radford's-Majid's theorem \cite{Majid8}
that any quantum braided group projection comes from cross product.

\interskip
Let $(A,{\cal R}_A)$ be a braided-quantum-group in $\cal C$
and $B$ a Hopf algebra in ${\cal C}_A^{{\cal R}_A}$.
Then one can use imbedding
 ${\cal C}_A^{{\cal R}_A}\rightarrow{\cal C}_A^A$
to construct cross-produt Hopf algebra $A\ltimes B$.
Combineing (\ref{CrossProd}) and (\ref{comodstr}) one can obtain expresion
for comultiplication in $A\ltimes B$:
\begin{equation}
\Delta_{A\ltimes B}\quad :=\quad
\matrix{\hstep\object{A}\Step\step\object{B}\step\cr
        \vvbox{\hbox{\hcd\step\hstep\cd}
	       \hbox{\id\step\id\hstep\dd\hhstep\ra\hstep\id}
	       \hhbox{\id\step\hru\hstep\id\Step\id\hstep\id}
	       \hbox{\id\step\hx\Step\id\hstep\id}
	       \hbox{\id\step\id\step\cu\hstep\id}}\cr
	\object{A}\step\object{B}\Step\object{A}\step\hstep\object{B}}
\label{comul}
\end{equation}

And let's try to define any $\cal R$-element ${\cal R}_H$
for $H=A\ltimes B$.
Condition
\begin{equation}
\vvbox{\hbox{\id\step\ro{{\cal R}_H}}
       \hbox{\ru\Step\id}}
\quad=\quad
\vvbox{\hbox{\id\step\Rb}
       \hbox{\ru\step\ra\step\id}
       \hbox{\Ru\Step\id\step\id}}
\end{equation}
prompts us the following formula
\begin{equation}
{\cal R}_{A\ltimes B}\quad :=\quad
\vvbox{\hbox{\Rb}
       \hbox{\id\step\hstep\ra\hstep\id}
       \hhbox{\id\step\cd\step\hstep\id\hstep\id}
       \hbox{\hx\step\id\step\hstep\id\hstep\id}
       \hbox{\id\step\ru\step\hstep\id\hstep\id}}
\label{Relement}
\end{equation}

According to lemma \ref{opqugr}
$(A_{\widetilde{\rm op}},{\cal R}_{A_{\widetilde{\rm op}}})$
is a quantum group in category $\overline{\cal C}$ and
$(B_{\widetilde{\rm op}},{\cal R}_{B_{\widetilde{\rm op}}})$
is a quantum group in
$\overline
  {\cal C}^{{\cal R}_{A_{\widetilde{\rm op}}}}_{A_{\widetilde{\rm op}}}$.
And one can define cross product Hopf algebra
$A_{\widetilde{\rm op}}\ltimes B_{\widetilde{\rm op}}$.
Multiplication coincide with one on $A\ltimes B$.
Comultiplication, which defined by the formula (\ref{comul}) applyed to data
as above, and $\cal R$-element (\ref{Relement}) take the forms:
\begin{equation}
\Delta_{A_{\widetilde{\rm op}}\ltimes B_{\widetilde{\rm op}}}:=
\matrix{\hstep\object{A}\Step\object{B}\Step\step\hstep\cr
\vvbox{\hbox{\obj{\widetilde\Delta}\hcd\step\hcd\obj{\widetilde\Delta}
             \step\rra}
       \hbox{\hxx\step\id\step\hx\step\dd}
       \hbox{\id\step\id\step\ru\step\ru}
       \hbox{\id\step\id\step\xx}
       \hbox{\id\step\id\step\id\hstep\rra\hhstep\d}
       \hbox{\id\step\id\step\id\hstep\xx\hstep\id}
       \hhbox{\id\step\id\step\hru\Step\id\hstep\id}
       \hbox{\id\step\hxx\Step\hstep\id\hstep\id}
       \hhbox{\id\step\id\step\d\Step\id\hstep\id}
       \hbox{\id\step\id\step\hstep\cu\hstep\id}}\cr
\object{A}\step\object{B}\Step\hstep\object{A}\step\hstep\object{B}\step}
\qquad\quad
{\cal R}_{A_{\widetilde{\rm op}}\ltimes B_{\widetilde{\rm op}}}:=\quad
\vvbox{\hbox{\rrb\step\rra}
       \hbox{\d\step\hx\step\dd}
       \hbox{\step\ru\step\ru}
       \hbox{\step\xx}
       \hbox{\dd\hstep\rra\hhstep\d}
       \hbox{\id\step\hstep\xx\hstep\id}
       \hhbox{\id\step\cd\obj{\widetilde\Delta}\step\hstep\id\hstep\id}
       \hbox{\id\step\hxx\step\hstep\id\hstep\id}
       \hbox{\hxx\step\id\step\hstep\id\hstep\id}
       \hbox{\id\step\ru\step\hstep\id\hstep\id}}
\end{equation}

\begin{theorem}
\begin{itemize}
\item
Hopf algebra $A\ltimes B$ is a quantum braided group in $\cal C$
with $\cal R$-element defined by (\ref{Relement}) and with the second
comultiplication
\begin{equation}
\widetilde\Delta_{A\ltimes B}:=
(\Delta_{A_{\widetilde{\rm op}}\ltimes B_{\widetilde{\rm op}}})^{\rm op}=
{}^{\cal C}\Psi_{A\otimes B,A\otimes B}\circ
\Delta_{A_{\widetilde{\rm op}}\ltimes B_{\widetilde{\rm op}}}
\label{secondcom}
\end{equation}
\item
Braided categories ${\cal C}^{{\cal R}_{A\ltimes B}}_{A\ltimes B}$ and
$({\cal C}^{{\cal R}_A}_A)^{{\cal R}_B}_B$ are isomorphic.
\end{itemize}
\label{quantumcross}
\end{theorem}

\begin{lemma}
\begin{equation}
{\cal R}_{A_{\widetilde{\rm op}}\ltimes B_{\widetilde{\rm op}}}=
{\cal R}_{(A\ltimes B)_{\widetilde{\rm op}}}
\end{equation}
\end{lemma}

\begin{proof}
Equalities on figure \ref{randinverse} show that
\begin{displaymath}
({}^{\cal C}\Psi_{A\otimes B,A\otimes B}\circ
{\cal R}_{A_{\widetilde{\rm op}}\ltimes B_{\widetilde{\rm op}}})
\;\cdot_{(A\ltimes B)\otimes(A\ltimes B)}\;
{\cal R}_{A\ltimes B}=
1_{A\ltimes B}\otimes 1_{A\ltimes B}
\end{displaymath}
Other side invertibility can be verifyed analogously.
\end{proof}

\begin{lemma}
$(A\ltimes B, {\cal R}_{A\ltimes B})$ is quasirtiangular bialgebra in
$\cal C$ with second comultiplication (\ref{secondcom}).
\end{lemma}

\begin{proof}
Second identity from (\ref{rmatrix}) for ${\cal R}_{A\ltimes B}$ is equivalent
first one for
${\cal R}_{A_{\widetilde{\rm op}}\ltimes B_{\widetilde{\rm op}}}$.
So it is enough to verify only the first
(See figures \ref{proofpairing1}--\ref{proofpairing2}).

Write briefly (\ref{adjoint}) in the form
$$\widetilde\Delta_A\cdot{\cal R}_A=
  {\cal R}_A\cdot\Delta_A            $$
$$\widetilde\Delta_B\cdot{\cal R}_B=
  {\cal R}_B\cdot\Delta_B            $$
In the last formula dot means multiplication in algebra
$B\underline\otimes B$ where tensor product is in category
${\cal C}_A^{{\cal R}_A}$.
Then after series of transformations one can verify that
\begin{equation}
\widetilde\Delta_{A\ltimes B}\cdot{\cal R}_{A\ltimes B}
\quad =\quad
\matrix{\hstep\object{A}\Step\step\object{B}\step\Step\cr
        \vvbox{\hbox{\Obj{\Delta_A\cdot{\cal R}_A}\hcd\step\hstep
                     \cd\Step\Obj{\Delta_B\cdot{\cal R}_B}}
	       \hbox{\id\step\id\hstep\dd\hhstep\ra\hstep\id}
	       \hhbox{\id\step\hru\hstep\id\Step\id\hstep\id}
	       \hbox{\id\step\hx\Step\id\hstep\id}
	       \hbox{\id\step\id\step\cu\hstep\id}}\cr
	\object{A}\step\object{B}\Step\object{A}\step\hstep\object{B}\Step}
\quad =\quad
{\cal R}_{A\ltimes B}\cdot\Delta_{A\ltimes B}
\end{equation}
\end{proof}

\begin{proof}(of the theorem \ref{quantumcross})
Moreover $(A\ltimes B, {\cal R}_{A\ltimes B})$ is quantum braided group
because both $A\ltimes B$ and
$A_{\widetilde{\rm op}}\ltimes B_{\widetilde{\rm op}}$ are Hopf algebras
according to the theorem \ref{CrossProduct}.

A proof of the second part is on the figure \ref{proofweakopposite}
where it is shown that any object of
$({\cal C}^{{\cal R}_A}_A)^{{\cal R}_B}_B$ belongs to
${\cal C}^{{\cal R}_{A\ltimes B}}_{A\ltimes B}$.
\end{proof}

As in the case of Hopf algebras (theorem \ref{transitivity})
quantum group cross product is transitive:

\begin{theorem}
Let $(A,{\cal R}_A)$ be a quantum group in $\cal C$,
$(B,{\cal R}_B)$ a quantum group in ${\cal C}^{{\cal R}_A}_A$,
$(C,{\cal R}_C)$ a quantum group in
${\cal C}^{{\cal R}_{A\ltimes B}}_{A\ltimes B}=
({\cal C}^{{\cal R}_A}_A)^{{\cal R}_B}_B$.
Then quantum groups $(A\ltimes B)\ltimes C$ and $A\ltimes (B\ltimes C)$
coincide.
\end{theorem}

\begin{proof}
It is a corollary of theorem \ref{transitivity}.
One needs only to verify that
${\cal R}_{(A\ltimes B)\ltimes C}={\cal R}_{A\ltimes (B\ltimes C)}$
using associativity of multiplication in $A\ltimes B\ltimes C$.
\end{proof}

\interskip
\begin{definition}
Let $(A,{\cal R}_A)$ and $(H,{\cal R}_H)$ be quantum groups in braided
category $\cal C$.
Pair of morphisms
$A\matrix{i_A\cr
	  \longrightarrow\cr
	  \longleftarrow\cr
	  p_A}              H$
is called {\em a quantum group projection} if
\begin{itemize}
\item
$i_A, p_A$ are Hopf algebra morphisms (in the both cases if we consider
$H$ and $A$ with first comultiplications
$\Delta_H,\Delta_A$
or with second ones
$\widetilde\Delta_H,\widetilde\Delta_A$) and
$p_A\circ i_A={\rm id}_A$.
Denote idempotent $i_A\circ p_A$ by $\Pi_A$.
\item
\begin{equation}
\vvbox{\hbox{\ra}}
\quad=\quad
\vvbox{\hbox{\rh}
       \hbox{\O{p_A}\Step\O{p_A}}}
\qquad\qquad
\vvbox{\hbox{\rh}
       \hbox{\O{\Pi_A}\Step\id}}
\quad=\quad
\vvbox{\hbox{\rh}
       \hbox{\O{\Pi_A}\Step\O{\Pi_A}}}
\quad=\quad
\vvbox{\hbox{\rh}
       \hbox{\id\Step\O{p_A}}}
\end{equation}
\item
The following idempotents coincide
\begin{equation}
\Pi_B\quad:=\quad
\vvbox{\hhbox{\krrr\cd\obj{\Delta_H}}
       \hbox{\O{p_A}\step\id}
       \hbox{\O{S_A}\step\id}
       \hbox{\O{i_A}\step\id}
       \hhbox{\krrr\cu}}
\quad=\quad
\vvbox{\hhbox{\krrr\cd\obj{\widetilde\Delta_H}}
       \hbox{\hxx}
       \hbox{\O{p_A}\step\id}
       \hbox{\O{\widetilde  S^{-1}_A}\step\id}
       \hbox{\O{i_A}\step\id}
       \hhbox{\krrr\cu}}
\label{idempotents}
\end{equation}
\end{itemize}
\end{definition}

\begin{remark}
If $A$ and $H$ are usual quantum groups with
$\widetilde\Delta=\Delta^{\rm op}$ and
$\widetilde S=S^{-1}$
then (\ref{idempotents}) is satisfyed automaticaly.
\end{remark}

\begin{theorem}
Let $\cal C$ be a braided category with split idempotents and
$(A,{\cal R}_A)
\matrix{i_a\cr\longrightarrow\cr\longleftarrow\cr p_A}
(H,{\cal R}_H)$
a quantum group projection in $\cal C$.
Then there exists quantum group $(B,{\cal R}_B)$ in category
${\cal C}^{{\cal R}_A}_A$ such that $(H,{\cal R}_H)$ is
a quantum group cross product $(A\ltimes B,{\cal R}_{A\ltimes B})$.
\end{theorem}

\begin{proof}
Let
$H
\matrix{p_B\cr\longrightarrow\cr\longleftarrow\cr i_B}
B$
split idempotent $\Pi_B$ in $\cal C$.
According to theorem \ref{Radford} Hopf algebra $H_{\Delta_A}$
(with first comultiplication $\Delta_A$) is cross product
$A_{\Delta_A}\ltimes B_{\Delta_B}$,
where $B$ is Hopf algebra in the category ${\cal C}_A^A$.
Identity
$$(p_B\otimes p_A)\circ ({\cal R}_H\cdot\Delta_H)\circ i_B=
(p_B\otimes p_A)\circ (\widetilde\Delta_H\cdot{\cal R}_H)\circ i_B$$
means exactly that coaction
$\beta:\,B\rightarrow B\otimes A$
has the form (\ref{comodstr}) (see figure \ref{proofcomodstr}).
So together with proposition \ref{qugrmod} this show that $B$ is a Hopf algebra
in ${\cal C}^{{\cal R}_A}_A$.
The same considerations applyed to the pair of quasitriangular Hopf algebras
$H_{\widetilde{\rm op}},A_{\widetilde{\rm op}}$ in category
$\overline{\cal C}$ show that Hopf algebra
$H_{{\widetilde\Delta_H}^{\rm op}}$
is cross product
$A_{{\widetilde\Delta_A}^{\rm op}}\ltimes B_{{\widetilde\Delta_B}^{\rm op}}$,
where $B_{{\widetilde\Delta_B}^{\rm op}}$ is a Hopf algebra in
$\overline{\cal C}
 ^{{\cal R}_{A^{\widetilde{\rm op}}}}
           _{A^{\widetilde{\rm op}}} $
and then (see lemma \ref{opqugr})
$B_{\widetilde\Delta_B}$ is a Hopf algebra in ${\cal C} ^{{\cal R}_A}_A$.

Define
\begin{equation}
{\cal R}_B:=(p_B\otimes p_B)\circ{\cal R}_H
\end{equation}
Figure \ref{rprojb} show that
\begin{equation}
({\rm id}\otimes p_B)\circ{\cal R}_H=
(i_B\otimes{\rm id})\circ{\cal R}_B
\label{B-proj}
\end{equation}
Applying $\;{\rm id}_H\otimes p_A\otimes p_B\;$ to the first identity from
(\ref{rmatrix}) for ${\cal R}_H$ and taking into account (\ref{B-proj})
we obtain that ${\cal R}_H$ is cross product '$\cal R$-element'
(\ref{Relement}).
Identity
\begin{equation}
(p_B\otimes p_B)\circ({\cal R}_H\cdot\Delta_H)\circ i_A=
(p_B\otimes p_B)\circ(\widetilde\Delta_H\cdot{\cal R}_H)\circ i_A
\end{equation}
means exactly that ${\cal R}_B$ is $A$-module morphism.
Identities (\ref{rmatrix}-\ref{adjoint}) for $(B,{\cal R}_B)$
follow directly from corresponding identities for $(H,{\cal R}_H)$.
\end{proof}

\interskip

\begin{figure}
\begin{displaymath}
\vvbox{\hhbox{\id\step\cd}
       \hbox{\hx\step\id}
       \hhbox{\id\step\ru}
       \hbox{\id\step\id\step\r}
       \hhbox{\id\step\ru\Step\id}
       \hbox{\hx\Step\dd}
       \hbox{\id\step\cu}}
\quad ={}\quad
\vvbox{\hhbox{\id\step\cd $\widetilde\Delta$ }
       \hbox{\ru\step\id\step\r}
       \hbox{\id\Step\hx\Step\id}
       \hbox{\Ru\step\cu}}
\quad ={}\quad
\vvbox{\hbox{\id\step\r\step\cd}
      \hbox{\ru\Step\hx\Step\id}
      \hbox{\id\Step\dd\step\id\Step\id}
      \hbox{\Ru\Step\cu}}
\end{displaymath}
\begin{quotation}
The first identity is equivalent to (\ref{opposite})
(since $\cal R$ is invertible in $H\otimes H$ (\ref{invertible})).
The second is a corollary of (\ref{adjoint}).
\end{quotation}
\caption
{$X\in{\rm Obj}({\cal C}^{\cal R}_H)$ iff
 $X^{\cal R}\in{\rm Obj}({\cal C}_H^H)$.}
\label{brm}
\end{figure}

\newfigure
\begin{figure}
\begin{displaymath}
\vvbox{\hbox{\id\step\id\hstep\step\r}
       \hhbox{\id\step\id\step\cd\step\hstep\id}
       \hbox{\id\step\hx\step\id\step\hstep\id}
       \hbox{\ru\step\ru\step\hstep\id}}
\quad ={}\quad
\vvbox{\hbox{\id\step\id\hstep\step\r}
       \hhbox{\id\step\id\step\obj{\widetilde\Delta}\cd\hstep\step\id}
       \hbox{\id\step\ru\step\id\hstep\step\id}
       \hbox{\id\step\xx\hstep\step\id}
       \hbox{\ru\Step\id\hstep\step\id}}
\quad ={}\quad
\vvbox{\hbox{\id\Step\Step\id\step\r}
       \hbox{\id\step\r\step\ru\Step\id}
       \hbox{\ru\Step\hx\step\Step\id}
       \hbox{\id\Step\step\id\step\d\Step\id}
       \hbox{\id\step\Step\id\Step\cu}}
\end{displaymath}
\begin{quotation}
The first equality is (\ref{opposite}). The second is (\ref{rmatrix}).
\end{quotation}
\caption
{$(X\otimes Y)^{\cal R}=X^{\cal R}\otimes  Y^{\cal R}$.}
\label{brmm}
\end{figure}

\newfigure
\begin{figure}
\begin{displaymath}
\vvbox{\hbox{\tS\step\id}
       \hbox{\hx}
       \hbox{\ru}}
\quad ={}\quad
\vvbox{\hbox{\hstep\hx}
         \hhbox{\dd\hstep\cd\obj{\widetilde\Delta}}
         \hbox{\id\step\tS\hstep\hcd}
         \hbox{\ru\hstep\hxx}
         \hhbox{\id\step\dd\step\id}
         \hbox{\ru\step\hstep\SS}
         \hhbox{\id\Step\dd}
         \hbox{\Ru}}
\quad=\quad
\vvbox{\hbox{\SS\step\id}
       \hbox{\hxx}
       \hbox{\ru}}
\end{displaymath}
\begin{quotation}
The first equality is the antipode axiom for $S^{-1}$.
The second use (\ref{opposite}) and antipode axiom for $\widetilde S$.
\end{quotation}
\caption
{Replacement $S^{-1}$ by $\widetilde S$.}
\label{proofchant}
\end{figure}

\newfigure
\begin{figure}
\begin{displaymath}
\matrix{\object{X}\step\object{X^\vee}\step\object{H}\cr
\vvbox{\hbox{\id\step\id\step\S}
       \hbox{\id\step\hx}
       \hbox{\ru\step\id}
       \hbox{\ev}}}
\quad ={}\quad
\vvbox{\hbox{\id\Step\Step\hstep\hx}
       \hhbox{\id\Step\Step\cd\hstep\d}
       \hbox{\id\Step\step\hddcd\step\id\step\id}
       \hbox{\id\step\hcoev\step\tSS\step\id\step\S\step\id}
       \hbox{\id\step\id\step\hx\step\id\step\id\step\id}
       \hbox{\id\step\hxx\step\ru\step\id\step\id}
       \hbox{\ru\step\id\step\Ru\dd}
       \hbox{\ev\step\ev}}
\quad ={}\quad
\matrix{\object{X}\step\object{X^\vee}\step\object{H}\cr
\vvbox{\hbox{\id\step\id\step\tSS}
       \hbox{\id\step\hxx}
       \hbox{\ru\step\id}
       \hbox{\ev}}}
\end{displaymath}
\begin{quotation}
The first equality is the antipode axiom.
The second use (\ref{opposite}) and antipode axiom for $\widetilde S^{-1}$.
\end{quotation}
\caption{Replacement $S$ by $\widetilde S^{-1}$ when right dual exists.}
\label{proofchantt}
\end{figure}

\newfigure
\begin{figure}
\begin{displaymath}
\vvbox{\hbox{\id\step\R}
       \hbox{\id\step\SS\step\hcoev\Step\id}
       \hbox{\id\step\hxx\step\id\Step\id}
       \hbox{\id\step\ru\step\id\Step\id}
       \hbox{\hev\Step\id\Step\id}}
\quad ={}\quad
\vvbox{\hbox{\id\step\R}
       \hbox{\id\step\tS\step\hcoev\Step\id}
       \hbox{\id\step\hx\step\id\Step\id}
       \hbox{\id\step\ru\step\id\Step\id}
       \hbox{\hev\Step\id\Step\id}}
\quad ={}\quad
\vvbox{\hbox{\id\Step\coev}
       \hbox{\id\step\dd\r\d}
       \hbox{\id\step\ru\Step\hx}
       \hbox{\hev\Step\step\id\step\S}}
\end{displaymath}
\begin{quotation}
First identity follows from (\ref{changeantipode}).
\end{quotation}
\caption{Comodule structures on $({}^\vee X)^{\cal R}$ and
         ${}^\vee (X^{\cal R})$ are the same.}
\label{proofld}
\end{figure}

\newfigure
\begin{figure}
\begin{displaymath}
\vvbox{\hbox{\tS\Step\O{u}}
       \hbox{\cu}}
\enspace :=\enspace
\vvbox{\hbox{\r\step\cd}
       \hbox{\id\Step\hx\step\cd}
       \hbox{\cu\step\id\step\id\Step\S}
       \hbox{\step\tS\Step\id\step\cu}
       \hbox{\step\cu\step\dd}
       \hbox{\Step\cu}}
\enspace =\enspace
\vvbox{\hbox{\Step\cd}
       \hbox{\step\dd\Step\d}
       \hbox{\Obj{\widetilde\Delta}\cd\step\r\d}
       \hbox{\id\Step\hx\Step\id\step\S}
       \hbox{\cu\step\cu\step\id}
       \hbox{\step\tS\Step\dd\step\dd}
       \hbox{\step\cu\step\dd}
       \hbox{\Step\cu}}
\enspace =\enspace
\vvbox{\hbox{\hstep\cd}
       \hbox{\obj{\widetilde\Delta}\hcd\step\hstep\d}
       \hbox{\tS\step\id\hstep\r\d}
       \hbox{\hcu\hstep\tS\Step\id\step\id}
       \hbox{\hstep\hx\Step\id\step\S}
       \hbox{\hstep\id\step\cu\step\id}
       \hbox{\hstep\cu\step\dd}
       \hbox{\step\hstep\cu}}
\enspace =:\enspace
\vvbox{\hbox{\O{u}\Step\S}
       \hbox{\cu}}
\end{displaymath}
\begin{quotation}
The second equality is ${\cal R}\cdot\Delta=\widetilde\Delta\cdot{\cal R}$.
\end{quotation}
\caption
{Two antipodes $S$ and $\widetilde S$ are adjoint:
 $\widetilde S\cdot u=u\cdot S$.}
\label{proofadjant}
\end{figure}

\newfigure
\thispagestyle{empty}
\addtocounter{page}{-1}
\begin{figure}
\begin{displaymath}
({}^{\cal C}\Psi_{A\otimes B,A\otimes B}\circ
{\cal R}_{A_{\widetilde{\rm op}}\ltimes B_{\widetilde{\rm op}}})
\cdot_{(A\ltimes B)\otimes(A\ltimes B)} {\cal R}_{A\ltimes B}=
\end{displaymath}
\begin{displaymath}
=\enspace
\vvbox{\hbox{\Step\Step\step\rrb\step\rra}
       \hbox{\Rb\step\id\Step\hx\step\dd}
       \hbox{\id\step\ra\step\id\step\Ru\step\ru}
       \hbox{\id\step\hdcd\step\id\step\id\step\id\step\rra\d}
       \hbox{\hx\step\id\step\id\step\id\step\hxx\step\cd\d}
       \hbox{\id\step\ru\step\id\step\hx\step\id\step\hd\step\hxx}
       \hbox{\id\step\id\Step\hx\step\hx\step\hcd\step\ru}
       \hbox{\id\step\id\step\hddcd\step\hx\step\hx\step\id\step\id}
       \hbox{\id\step\hx\step\id\step\id\step\hcu\step\ru\step\id}
       \hbox{\hcu\step\ru\step\id\step\hstep\id\step\hstep\cu}
       \hbox{\hstep\id\step\hstep\cu\step\hstep\id\Step\hstep\id}}
\enspace =\enspace
\vvbox{\hbox{\Step\step\rrb\step\rra}
       \hbox{\rb\step\id\Step\hx\step\dd}
       \hbox{\id\Step\id\step\id\Ru\step\ru}
       \hbox{\id\Step\id\step\id\step\rra\d}
       \hbox{\id\Step\id\step\hxx\Step\id\step\id}
       \hbox{\id\Step\hx\step\id\Step\hxx}
       \hbox{\id\step\hddcd\step\hx\Step\id\step\id}
       \hbox{\hx\step\id\step\id\step\cu\dd}
       \hbox{\id\step\ru\step\id\Step\ru}
       \hbox{\id\step\cu\step\unit\step\id}}
=\enspace
\vvbox{\hbox{\rb\Step\step\rrb}
       \hbox{\id\Step\id\Step\dd\ra\d}
       \hbox{\id\Step\d\step\ru\step\dd\step\id}
       \hbox{\id\Step\step\hx\step\dd\step\dd}
       \hbox{\id\Step\dd\step\ru\step\dd}
       \hbox{\cu\rra\cu}
       \hbox{\step\id\step\hdcd\step\id\step\id}
       \hbox{\step\hx\step\id\step\hxx}
       \hbox{\step\id\step\ru\hstep\unit\hstep\ru}}
\end{displaymath}
\begin{quotation}
The second equality is 'the reduction' of ${\cal R}_A$ and ${\cal R}^{-1}_A$.
The second and the third equalities both use that
$B$ is object in ${\cal C}^{{\cal R}_A}_A$.
It is easy to see that the last diagram represent identical map.
\end{quotation}
\caption{${\cal R}_{A_{\widetilde{\rm op}}\ltimes B_{\widetilde{\rm op}}}=
          {\cal R}_{(A\ltimes B)_{\widetilde{\rm op}}}$}
\label{randinverse}
\end{figure}

\newfigure
\begin{figure}
\begin{displaymath}
({\rm id}\otimes\Delta_{A\ltimes B}){\cal R}_{A\ltimes B}:=
\end{displaymath}
\begin{displaymath}
\matrix{\vvbox{\hbox{\step\Rb}
               \hbox{\dd\hstep\ra\step\hstep\d}
               \hbox{\id\step\hcd\step\hcd\step\cd}
               \hbox{\hx\step\id\step\id\step\id\step\id\hstep\ra\hhstep\d}
               \hhbox{\id\step\ru\step\id\step\id\step\hru\step\hstep\dd
                                                               \hstep\id}
               \hbox{\id\step\id\Step\id\step\hx\step\dd\step\id}
               \hhbox{\id\step\id\Step\id\step\id\step\cu\Step\id}}\cr
	\object{A}\step\object{B}\Step\object{A}\step\object{B}\step
		  \object{A}\Step\step\object{B}}
\quad =\quad
\vvbox{\hbox{\Step\Rb}
       \hbox{\step\dd\Rb\d}
       \hbox{\dd\dd\Ra\d\d}
       \hbox{\id\step\id\step\id\step\ra\step\id\step\id\step\d}
       \hhbox{\id\step\id\step\cu\Step\id\step\id\step\id\Step\d}
       \hhbox{\id\step\id\step\cd\Step\id\step\id\step\id\Step\hstep\d}
       \hbox{\id\step\hx\step\id\Step\id\step\id\step\id\step\ra\d}
       \hbox{\hx\step\id\step\id\Step\id\step\id\step\ru\Step\id\step\id}
       \hbox{\id\step\hcu\step\id\Step\id\step\hx\Step\dd\step\id}
       \hhbox{\id\step\hstep\d\step\id\Step\id\step\id\step\d\step\dd\Step\id}
       \hbox{\id\Step\ru\Step\id\step\id\step\hstep\hcu\Step\hstep\id}}
\quad ={}
\end{displaymath}
\begin{quotation}
The second equality is the first axiom from (\ref{rmatrix}) for
${\cal R}_A$ and ${\cal R}_B$.
The third use bialgebra axiom for $A$ and that $B$ is $A$-module algebra.
To prove the 4th we use that $\widetilde\Delta_A$ is weak opposite
with respect to $B$ and then (\ref{rmatrix}-\ref{adjoint}) for ${\cal R}_A$.
And the 5th use that ${\cal R}_B$ is $A$-module morphism.
\end{quotation}
\caption
      {$({\rm id}\otimes\Delta_{A\ltimes B})\circ{\cal R}_{A\ltimes B}=
        ({\cal R}_{A\ltimes B})_{13}\cdot({\cal R}_{A\ltimes B})_{12}$
      (part 1)}
\label{proofpairing1}
\end{figure}

\newfigure
\begin{figure}
\begin{displaymath}
\kern -2\unitlength
=\enspace
\vvbox{\hbox{\Step\step\Rb}
       \hbox{\Step\dd\hhstep\rb\step\ra\hhstep\d}
       \hbox{\step\dd\hhstep\dd\Step\ru\ra\d\hhstep\d}
       \hbox{\dd\hstep\id\Step\step\hxx\Step\d\d\hhstep\d}
       \hbox{\id\step\hstep\id\Step\dd\step\d\Step\hcu\hstep\id}
       \hhbox{\id\step\hstep\d\step\cd\Step\hstep\d\Step\id\step\id}
       \hbox{\id\Step\hx\step\id\hstep\ra\hstep\d\step\id\step\id}
       \hhbox{\d\step\cd\hstep\ru\cd\step\hstep\id\step\hstep\id\step\id
                                                                \step\id}
       \hbox{\hstep\hx\step\id\hstep\hx\step\id\step\hstep\id
					       \step\hstep\id\step\id\step\id}
       \hhbox{\hstep\id\step\ru\cd\hstep\ru
                             \step\hstep\id\step\hstep\id\step\id\step\id}
       \hbox{\hstep\id\step\hx\step\id\hstep\id
                             \Step\hstep\id\step\hstep\id\step\id\step\id}
       \hhbox{\hstep\cu\step\ru\hstep\id
                             \Step\hstep\id\step\hstep\id\step\id\step\id}
       \hhbox{\step\id\step\hstep\id\step\dd
                             \Step\hstep\id\step\hstep\id\step\id\step\id}
       \hhbox{\step\id\step\hstep\cu
                             \Step\step\id\step\hstep\id\step\id\step\id}}
\enspace=\enspace
\vvbox{\hbox{\Step\Rb}
       \hbox{\step\dd\Step\hstep\ra\hhstep\d}
       \hbox{\dd\rb\step\hcd\step\hstep\d\hhstep\d}
       \hbox{\id\step\id\Step\hx\step\id\hstep\ra\d\hhstep\d}
       \hhbox{\id\step\Ru\step\ru\dd\Step\cu\hstep\id}
       \hbox{\id\step\id\Step\step\hxx\Step\step\id\step\id}
       \hbox{\id\step\d\step\dd\step\d\Step\id\step\id}
       \hbox{\id\Step\hxx\step\hstep\ra\hhstep\d\step\id\step\id}
       \hhbox{\d\step\cd\hstep\id\step\cd\step\hstep
                            \id\hstep\id\step\id\step\id}
       \hbox{\hstep\hx\step\id\hstep\hx\step\id\step\hstep
                            \id\hstep\id\step\id\step\id}
       \hhbox{\hstep\id\step\ru\cd\hstep\ru\step\hstep
                            \id\hstep\id\step\id\step\id}
       \hbox{\hstep\id\step\hx\step\id\hstep\id\Step\hstep
                            \id\hstep\id\step\id\step\id}
       \hhbox{\hstep\cu\step\ru\hstep\id\Step\hstep
                            \id\hstep\id\step\id\step\id}
       \hhbox{\step\id\step\hstep\id\step\dd\Step\hstep
                            \id\hstep\id\step\id\step\id}
       \hhbox{\step\id\step\hstep\cu\Step\step
                            \id\hstep\id\step\id\step\id}}
\enspace=\enspace
\vvbox{\hbox{\Step\Rb}
       \hbox{\step\dd\hstep\Ra\hhstep\d}
       \hbox{\dd\hstep\dd\step\rb\step\d\hhstep\d}
       \hbox{\id\step\hstep\id\step\dd\hstep\ra\hhstep\d\step\id\hstep\id}
       \hhbox{\id\step\cd\hstep\id\step\cd\step\hstep
			   \id\hstep\id\step\id\hstep\id}
       \hbox{\hx\step\id\hstep\hx\step\id\step\hstep
			   \id\hstep\id\step\id\hstep\id}
       \hhbox{\id\step\ru\cd\hstep\ru\step\hstep
			   \id\hstep\id\step\id\hstep\id}
       \hbox{\id\step\hx\step\id\hstep\id\Step\hstep
			   \id\hstep\id\step\id\hstep\id}
      \hhbox{\cu\step\ru\hstep\id\Step\hstep
			   \id\hstep\id\step\id\hstep\id}
      \hhbox{\hstep\id\step\hstep\id\step\dd\Step\hstep
			   \id\hstep\id\step\id\hstep\id}
      \hhbox{\hstep\id\step\hstep\cu\Step\step
			   \id\hstep\id\step\id\hstep\id}}
\enspace=
\end{displaymath}
\begin{displaymath}
=:({\cal R}_{A\ltimes B})_{13}\cdot({\cal R}_{A\ltimes B})_{12}
\end{displaymath}
\caption
      {$({\rm id}\otimes\Delta_{A\ltimes B})\circ{\cal R}_{A\ltimes B}=
        ({\cal R}_{A\ltimes B})_{13}\cdot({\cal R}_{A\ltimes B})_{12}$
      (part 2)}
\label{proofpairing2}
\end{figure}

\newfigure
\thispagestyle{empty}
\addtocounter{page}{-1}
\begin{figure}
\begin{displaymath}
\matrix{\object{X}\step\hstep\object{A\ltimes B}\hstep\cr
\vvbox{\hhbox{\id\step\cd}
       \hbox{\hx\step\id}
       \hbox{\id\step\ru}}}
\quad:=
\end{displaymath}
\begin{displaymath}
:=\quad
\vvbox{\hhbox{\id\step\cd\step\Step\id}
       \hbox{\hx\step\id\Step\cd}
       \hbox{\id\step\ru\step\dd\ra\d}
       \hbox{\id\step\d\step\ru\step\dd\step\id}
       \hbox{\id\Step\hx\step\dd\step\dd}
       \hbox{\id\step\dd\step\ru\step\dd}
       \hbox{\id\step\id\Step\Ru}}
\quad=\quad
\vvbox{\hhbox{\id\step\obj{\widetilde\Delta_A}\cd\step
                        \cd\obj{\widetilde\Delta_B}}
       \hbox{\ru\step\id\step\id\step\id}
       \hbox{\xx\step\id\step\id\step\rra}
       \hbox{\id\Step\ru\step\hx\Step\id}
       \hbox{\id\Step\Ru\step\Ru}
       \hbox{\id\Step\d\step\dd}
       \hbox{\id\Step\step\hxx}}
\quad=\quad
\vvbox{\hbox{\id\step\Obj{\widetilde\Delta_A}\hcd\step
             \cd\Obj{\widetilde\Delta_B}\step\rra}
       \hbox{\ru\step\id\step\id\Step\hx\Step\id}
       \hbox{\xx\step\Ru\step\Ru}
       \hbox{\id\Step\id\step\id\step\rra\d}
       \hbox{\id\Step\id\step\hxx\Step\hxx}
       \hbox{\id\Step\ru\step\id\Step\ru}
       \hbox{\id\Step\Ru\step\dd}
       \hbox{\id\Step\d\step\dd}
       \hbox{\id\Step\step\hxx}}
\quad =:
\end{displaymath}
\begin{displaymath}
=:\quad
\matrix{\object{X}\step\hstep\object{A\ltimes B}\hstep\cr
        \vvbox{\hhbox{\id\step\cd\obj{\widetilde\Delta}}
               \hbox{\ru\step\id}
               \hbox{\xx}}}
\end{displaymath}
\begin{quotation}
The second equality use that $X$ belongs to ${\cal C}^{{\cal R}_A}_A$ and
$({\cal C}^{{\cal R}_A}_A)^{{\cal R}_B}_B$.
The third use that $B$-action is $A$-module map.
\end{quotation}
\caption
        {Braided categories ${\cal C}^{{\cal R}_{A\ltimes B}}_{A\ltimes B}$
         and $({\cal C}^{{\cal R}_A}_A)^{{\cal R}_B}_B$ coincide.}
\label{proofweakopposite}
\end{figure}

\newfigure
\thispagestyle{empty}
\addtocounter{page}{-1}
\begin{figure}
\begin{displaymath}
\vvbox{\hbox{\Step\Step\O{i_B}}
       \hbox{\rh\step\cd\Obj{\Delta_H}}
       \hbox{\id\Step\hx\Step\id}
       \hbox{\cu\step\cu}
       \hbox{\step\O{p_B}\Step\step\O{p_A}}}
\quad=\quad
\vvbox{\hbox{\rh\Step\O{i_B}}
       \hbox{\id\Step\O{p_A}\step\cd}
       \hbox{\id\Step\hx\Step\O{p_A}}
       \hbox{\cu\step\cu}
       \hbox{\step\O{p_B}\Step\step\id}}
\quad=\quad
\vvbox{\hbox{\ra\step\rd\Obj{\beta}}
       \hbox{\O{i_A}\Step\hx\step\id}
       \hbox{\cu\step\hcu}
       \hbox{\step\O{p_B}\Step\hstep\id}}
\quad=\quad
\vvbox{\hbox{\rd\Obj{\beta}}}
\end{displaymath}
\begin{displaymath}
\vvbox{\hbox{\step\O{i_B}}
       \hbox{\cd\Obj{\widetilde\Delta_H}\step\rh}
       \hbox{\id\Step\hx\Step\id}
       \hbox{\cu\step\cu}
       \hbox{\step\O{p_B}\Step\step\O{p_A}}}
\quad=\quad
\vvbox{\hbox{\step\O{i_B}}
       \hbox{\cd\Obj{\widetilde\Delta_H}\step\rh}
       \hbox{\id\Step\O{p_A}\step\id\Step\O{p_A}}
       \hbox{\id\Step\hx\Step\id}
       \hbox{\cu\step\cu}
       \hbox{\step\O{p_B}\Step\step\id}}
\quad=\quad
\vvbox{\hbox{\id\step\ra}
       \hbox{\O{i_B}\step\O{i_A}\Step\id}
       \hbox{\hcu\Step\id}
       \hbox{\hstep\O{p_B}\Step\hstep\id}}
\quad=\quad
\vvbox{\hbox{\id\hstep\ra}
       \hhbox{\hru\Step\id}}
\end{displaymath}
\begin{quotation}
Here we use identities from the proof of the theorem \ref{CrossProduct}
and identity
$$({\rm id}\otimes p_A)\circ\widetilde\Delta_H\circ i_B=i_B\otimes{\rm id}_A$$
which is left analog of
$$(p_A\otimes{\rm id})\circ\Delta_H\circ i_B={\rm id}_A\otimes i_B$$
\end{quotation}
\caption{Comodule structure arising from quantum group projection.}
\label{proofcomodstr}
\end{figure}

\newfigure
\begin{figure}
\begin{displaymath}
\vvbox{\hbox{\rh}
       \hbox{\O{\Pi_B}\Step\O{\Pi_B}}}
\quad=\quad
\vvbox{\hbox{\hstep\rh}
       \hhbox{\cd\obj{\widetilde\Delta_H}\step\hstep\id}
       \hbox{\hxx\step\hstep\O{\Pi_B}}
       \hbox{\O{p_A}\step\id\step\hstep\id}
       \hbox{\tSS\step\id\step\hstep\id}
       \hbox{\O{i_A}\step\id\step\hstep\id}
       \hhbox{\cu\step\hstep\id}}
\quad=\quad
\vvbox{\hbox{\Rh}
       \hbox{\id\step\rh\step\id}
       \hbox{\hxx\Step\hcu}
       \hbox{\O{p_A}\step\id\Step\hstep\id}
       \hbox{\tSS\step\id\Step\hstep\O{\Pi_B}}
       \hbox{\O{i_A}\step\id\Step\hstep\id}
       \hhbox{\cu\Step\hstep\id}}
\quad=
\end{displaymath}
\begin{displaymath}
=\quad
\vvbox{\hbox{\Rh}
       \hbox{\id\step\rh\step\id}
       \hbox{\hxx\Step\O{i_A}\step\id}
       \hbox{\tSS\step\id\Step\hcu}
       \hbox{\O{i_A}\step\id\Step\hstep\O{\Pi_B}}
       \hhbox{\cu\Step\hstep\id}}
\quad=\quad
\vvbox{\hbox{\Rh}
       \hbox{\id\step\ra\step\id}
       \hbox{\hxx\Step\counit\step\id}
       \hbox{\tSS\step\id\Step\step\O{\Pi_B}}
       \hbox{\O{i_A}\step\id\Step\step\id}
       \hhbox{\cu\Step\step\id}}
\quad=\quad
\vvbox{\hbox{\rh}
       \hbox{\id\Step\O{\Pi_B}}}
\end{displaymath}
\begin{quotation}
The 4th equality use that
$\Pi_B\circ\cdot_H\circ(i_A\otimes{\rm id}_H)=\epsilon_A\otimes\Pi_B$.
\end{quotation}
\caption{$(\Pi_B\otimes\Pi_B)\circ{\cal R}_H=
          ({\rm id}\otimes\Pi_B)\circ{\cal R}_B$.}
\label{rprojb}
\end{figure}

\end{document}